\documentclass[manuscript]{acmart} 
\AtBeginDocument{%
  \providecommand\BibTeX{{%
    \normalfont B\kern-0.5em{\scshape i\kern-0.25em b}\kern-0.8em\TeX}}}

\acmBooktitle{Woodstock '18: ACM Symposium on Neural Gaze Detection,
 June 03--05, 2018, Woodstock, NY} 
\acmPrice{15.00}
\acmISBN{978-1-4503-XXXX-X/18/06}

\usepackage{xcolor}
\begin{document}

\title{User Experiences with Third-Party SIM Cards and ID Registration in Kenya and Tanzania}


\author{Edith Luhanga}
\affiliation{%
  \institution{Carnegie Mellon University-Africa}
  \country{Rwanda}
}

\author{Karen Sowon}
\affiliation{%
  \institution{Carnegie Mellon University}
  \country{USA}
  }

\author{Lorrie Faith Cranor}
\affiliation{%
  \institution{Carnegie Mellon University}
  \country{USA}
}

\author{Giulia Fanti}
\affiliation{%
 \institution{Carnegie Mellon University}
 \country{USA}
 }

\author{Conrad Tucker}
\affiliation{%
  \institution{Carnegie Mellon University}
  \country{USA}
  }

\author{Assane Gueye}
\affiliation{%
  \institution{Carnegie Mellon University-Africa}
  \country{Rwanda}
  }

\renewcommand{\shortauthors}{Luhanga and Sowon, et al.}

\begin{abstract}
  Mobile money services in Sub-Saharan Africa (SSA) have increased access to financial services. To ensure proper identification of users, countries have put in place ``Know-Your-Customer'' (KYC) measures such as SIM registration using an official identification. However, half of the 850 million people without IDs globally live in SSA, and the use of SIM cards registered in another person’s name (third-party SIM) is prevalent. In this study, we explore challenges that contribute to and arise from the use of third-party SIM cards. We interviewed 36 participants in Kenya and Tanzania. Our results highlight great strides in ID accessibility, but also highlight numerous institutional and social factors that contribute to the use of third-party SIM cards. While privacy concerns contribute to the use of third-party SIM cards, third-party SIM card users are exposed to significant security and privacy risks, including scams, financial loss, and wrongful arrest.
\end{abstract}





\maketitle

\section{Introduction}
\label{Sec:Introduction}

Sub Saharan Africa (SSA) had 515 million mobile subscribers (46\% of the population) at the end of 2021, with most of these using prepaid SIM cards. The number is expected to rise by 100 million by 2025 \cite{okeleke2019mobile}. Increased adoption of mobile services has led to increased use of digital financial services. Between 2011 and 2021, financial account ownership in SSA rose from 23\% to 53\%, largely because of increased use of mobile money \cite{demirgucc2022global}. 

Mobile money is a service where mobile network operators (MNOs) provide financial products such as transfers (peer-to-peer, person-to-government, person-to-business and vice versa), savings, microloans, and insurance through a user's phone. The services can be accessed through a USSD menu or through a smartphone app. To open a mobile money account, users first register a SIM card and then activate the mobile money account. Once registered for mobile money, the mobile phone number serves as the mobile money account number and the name registered to the SIM card serves as the account name.  

Rising crime and fraud perpetrated through mobile phones has led to widespread implementation of know-your-customer (KYC) measures to aid proper identification of mobile users. Globally, 157 countries, including almost all countries to SSA, require \emph{SIM card registration} using official IDs --- usually a national ID \cite{mobileidentity}. Several countries also collect biometrics during SIM registration. The number of SIM cards that can be registered under one ID is often limited. Tanzania allows 1 SIM card per network for voice, SMS, and data use and 4 SIM cards for machine-to-machine communication \cite{MinorSIMTz} while Kenya allows 10 SIM cards per ID \cite{KenyaSIMReg}.  

Despite the requirement to register SIM cards under an official ID, in practice, many users register  SIM cards under another person's ID; this practice is called \emph{third-party SIM usage.} In 7 low-and middle-income countries (LMICs), 3 of which are countries in SSA \footnote{The 3 SSA countries included in the survey were Kenya, Mozambique, and Nigeria.}, 18\% of SIM cards are registered under a third-party ID. Most users of third-party SIM cards in these countries do not have an ID; however, 17\% of people with some form of official ID also use the third-party SIMs \cite{mobileidentity}. 

Accessing mobile money with a third-party SIM card is akin to using a bank account registered under another person's name. In some countries, including Ghana and Kenya, in addition to being required to register SIM cards with an ID, users have to present the ID used to register the mobile money account whenever they withdraw money \cite{statemobilemoney}. Not having an ID therefore limits access to services, including financial services like mobile money. Users who lose or damage their SIM card may also lose access to their savings as the original ID, and sometimes biometrics, are needed to replace a SIM card. Additionally, international regulations require all financial providers, including mobile money providers, to conduct periodic reverification of customer identities and risk profiles \cite{kipkemboi2019overcoming}. Not being able to meet the requirements for reverification can lead to the SIM card being deactivated. While the consequences of third-party SIM card use can be significant, there is a paucity of research  exploring the risks and challenges that third-party SIM card users experience and how they navigate them. We also conjectured that the lack of IDs was likely associated to the processes of ID acquisition thus necessitating an in-depth analysis of these processes from a user perspective.

We therefore present the results of an exploratory qualitative study to answer the following research questions: 

\begin{itemize}
    \item RQ1: What privacy and security implications emerge from the use of third-party SIM cards? 
    \item RQ2: What factors contribute to the use of third-party SIM cards?
    \item RQ3: What are people's overall perceptions on mandatory SIM card registration?  
\end{itemize}

We conducted qualitative interviews with 72 people in two SSA countries with notably high adoption of mobile money, Kenya and Tanzania~\cite{statemobilemoney}. East Africa, which is a smaller region of SSA that includes Kenya and Tanzania, continues to have the highest number of registered mobile money accounts in the world (296 million in 2021) \cite{statemobilemoney}. 

Our findings reveal important insights regarding the use of third-party SIM cards and ID registration experiences: 

\begin{itemize}
    \item Our first finding shows that most people use the IDs of social connections like family and friends to register third-party SIMs. However, some do the registration without the ID owner's consent. MNO agents---people or businesses contracted to register SIM cards and facilitate mobile money transactions---also sell pre-registered SIM cards using previous clients' ID details, without their consent. Furthermore, ID owners register third-party SIMs for others despite having concerns over potential misuse.
    
    \item Second, third-party SIM card usage exposes people to several challenges and risks including those related to privacy and security. The users experience challenges with SIM registration and replacement procedures, difficulties in using mobile money, and workarounds employed reduce privacy and security over their mobile money account and transactions and can expose both the third-party SIM users and ID owners to the possibility of being implicated in crimes. 
    
    \item Third, people usually get third-party SIMs because of difficulties with getting an ID. Other non-ID related factors like registration misconceptions, and the desire for anonymity among users consequently motivates the use of third-party SIM cards. We further observe that the challenges in ID registration are not just experienced once---they recur at different stages of the ID registration process and result in most people needing more than one attempt to successfully apply for an ID.  
    
    \item Fourth, overall, many users support SIM card registration as a means for curbing rising levels of mobile money fraud. Among those who disapprove of the regulations are those who are motivated by a desire for anonymity and privacy. 
\end{itemize}   

Based on the insights from the study, we offer 3 recommendations that address the issues that the study elicited. First, to curb non consensual SIM card registration, we propose researching technologies that can allow self-registration and bypassing of agents, e.g., eSIMs, or the use 2-step verification during registration. Second, we discuss the need for further research on privacy-preserving mobile money interfaces to address users' desire for anonymity when transacting with strangers. Third, we recommend solutions to challenging ID user journeys. These include usability studies on online ID application systems, alternative identity proofing methods suited for the African context, expansion of expand ID access measurements to measure the number of people who get an ID at the first attempt and within a reasonable waiting time

\section{Background}

\begin{figure*}
\includegraphics[scale=0.5]{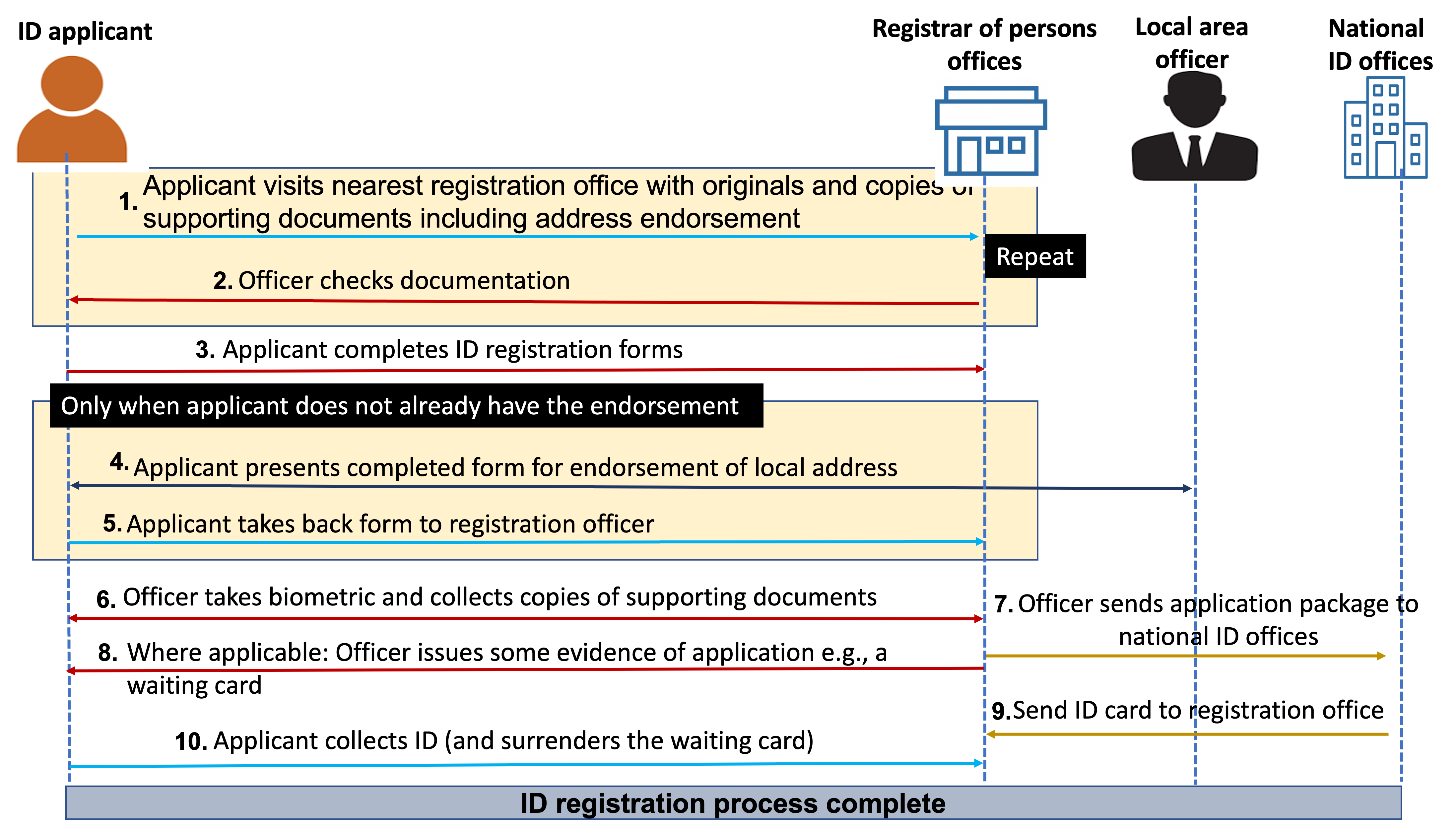}
\caption{A generalized illustration of the ID registration process
}
\Description{Flowchart depicting a generalized ID registration process in Kenya and Tanzania, which consists of 10 steps. The start state is the applicant visiting the registration office with the required supporting documents. Step 2 is officers checking the document. Step 3 is the applicant completing the ID registration form. If the applicant does not have an endorsement letter from local government officials, they complete steps 4 and 5 which are to present the completed form for endorsement to the local government officials and to return the form to the registration office. For all applicants, after completing the ID registration form, the officer takes biometrics and copies of the supporting documents in step 6. Step 7 is the sending of the application packet to the national ID offices, as the process is centralized. Step 8 is the applicant being issued proof of application e.g., a waiting card - if applicable. Step 9 is the processed ID card being sent to the local registration office and step 10 is the collection of the ID card by the applicant.}
\label{figIDProcess}
\end{figure*}

Figure \ref{figIDProcess} provides a generalized illustration of the ID application process in Kenya and Tanzania. The processing of IDs is centralized in both countries. Qualifying individuals can however apply for their IDs at specified local government offices within their community. Applicants are required to present themselves and required documentation, which is checked and approved by the local government officer. The documentation should provide proof of citizenship, local address, and age, and will usually be a combination of the following: birth certificate, clinic card\footnote{A newborn's medical record that helps parents and medical professionals keep track of immunizations, development and overall growth and health in their early life}, affidavit of birth\footnote{A sworn statement of facts around an individual's birth including birth date, location, and the name of each parent where a person does not have a birth certificate--- which is an official document that's issued by the government}, passport, driver's licence, school leaving certificate, an endorsement from a local official, land title
deed, and parents’ IDs (or death certificates and burial permits if deceased).

Once the officer is satisfied with the documentation, biometrics will usually be captured before the application package is sent to the national ID offices for the physical ID card to be processed. The officer provides the applicant a tentative date when they can expect to collect their ID from the local office after it has been sent by the national office. The following sections  provide further details of the ID application processes in the two countries.  
\subsection{Kenya}
Kenya's civil registration is well established, dating back to its pre-independence \cite{gelb2016identification}. The ID is required to access virtually all government services, file taxes, register mobile SIM cards, receive cash transfers, enroll in pension and social insurance programs, open banking and mobile money accounts, and to register to vote \cite{gelb2016identification}. Kenyan IDs are issued to qualifying individuals by the National Registration Bureau (NRB) and only expire upon the death  \cite{gelb2016identification}. 
The law gives the registration officer power to demand proof of information \cite{kenyalaw_2019}.
On paper, the registration process from initiation to getting the ID should take 30 days.  

\begin{itemize}
    \item \textbf{Requirements and eligibility:} All Kenyan citizens over the age of 18 years qualify for an ID and are required to begin the registration process within 90 days of attaining this age. Foreigners who remain in the country for more than 90 days are also required to get an alien registration card \cite{kenyalaw_2019}. Once supporting documents are deemed satisfactory, the local government office collects ink fingerprints and a facial photo. 
    \item \textbf{ID collection:} Applicants collect their IDs at the local government offices (Chief's or district commissioner's office) where the applicant was registered, after presenting the waiting card issued. According to grey sources\footnote{https://www.how.co.ke/how-to-track-the-status-of-your-id-application-by-sms/}\footnote{https://kenyayote.com/ask/10232/how-do-i-check-if-my-id-card-is-ready-for-collection}., applicants can check whether their IDs are ready for collection via SMS (texting the waiting card serial number to a shortcode) or via an online portal. 
    \item \textbf{Fees and costs:} The first application for a national ID is free, with replacements due to loss or damage costing Kshs. 100 (\$0.75)  and change of details costing Kshs. 300 to Kshs. 1,000 (\$2.24 to \$7.46), depending on the request\cite{KenyaIDEvolution}.\footnote{1 USD is equivalent to Kshs. 134.13 in April 2022} Alien IDs cost KShs. 1,000 per year (\$7.46) \cite{ForeignIDCost}. 
\end{itemize}

In 2019, Kenya rolled out a National Integrated Identity Management System, also known as ``Huduma namba'' that was to combine all existing user identities into one ``single source of truth.'' By September 2021, 10 million cards had been issued \cite{HudumaNumberControversy}. However, the rollout was halted in October 2022 over concerns about data privacy and exclusion of those who do not have a national ID \cite{HudumaNambaRuling}. 
\begin{figure*}[hbt!]
\includegraphics[width=\linewidth,scale=0.5]{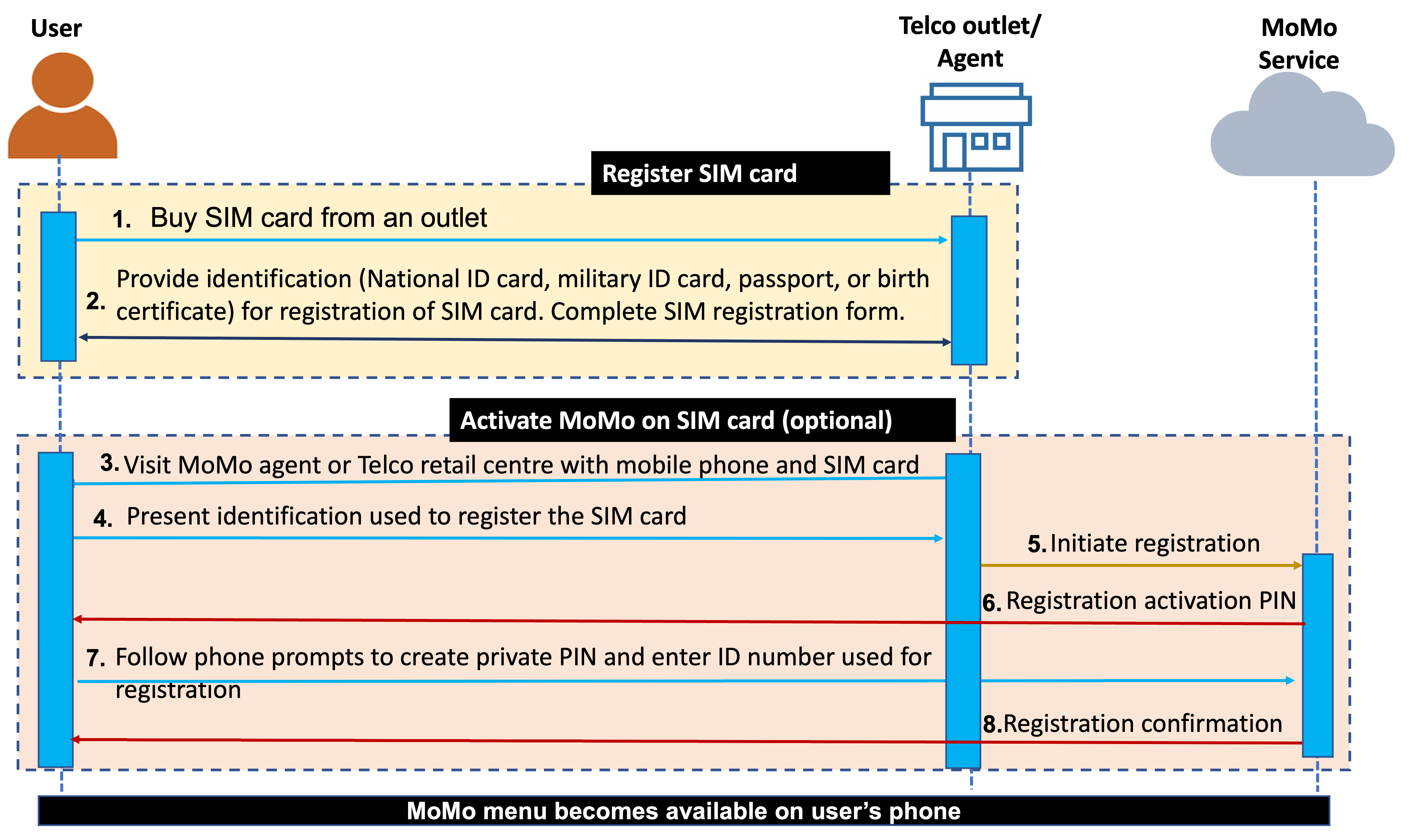}
\caption{SIM card and MoMo registration}
\label{figSIMRegistration}
\Description{Flowchart depicting the SIM card and mobile money registration process, which consists of 8 steps. The first step is buying a SIM card from an outlet. Step 2 is providing the identification e.g., national ID card, and in some cases the biometrics. A SIM application form is sometimes also filled. Steps 3 to 8 are only completed by those who want to activate their mobile money account. Step 3 is to visit the outlet with the registered SIM card. Step 4 is presenting the ID used to register the SIM card. Step 5 is initiating the registration by connecting to the mobile money service portal. Step 6 is choosing and inputting the registration activation PIN. Step 7 is following phone prompts to create a private PIN and to input the ID number used to register the SIM card. The final step, step 8, is receiving a mobile money registration confirmation.}
\end{figure*}

\subsection{Tanzania}
The government of Tanzania passed the ``Identification of Persons Act'' in 1986 but legal registration only began in 2013 \cite{TZNIDACivic}. The authority responsible for issuing IDs is the National Identification Authority (NIDA). All citizens aged 18 years and above, and all legal aliens (not under a tourist visa) are eligible for a national ID. Applicants follow the following process \cite{NIDAProcess}:
\vspace{-\topsep}
\begin{itemize}
    \item \textbf{Biographic information and biometrics:} The applicant submits an application form and supporting documents at the NIDA offices where a facial photo, fingerprints and a digital signature are collected.
    \item \textbf{Collection:} The applicant collects the printed ID from the local district office or another specified location.  
    \item \textbf{Fees and costs:} IDs are issued free-of-charge to citizens on first application while foreigner IDs cost between \$20 and \$100, depending on the type of visa held \cite{NIDACost}. All IDs expire after 10 years. Replacement IDs (due to loss, damage, or expiration) cost Tshs. 20,000 (approximately \$8.51) for citizens and the same as first application costs for foreigners. Additional replacements during the 10-year period incur successively higher costs with each new application \cite{NIDACost}. 
\end{itemize}

As of December 2022, 23 million people (around 71.7\% of adults) had been registered, with 19.845 million ID numbers and 11.187 million ID cards created \cite{NIDAonline}. In December 2022, NIDA introduced an online application portal to ease congestion at NIDA offices and to speed up the data entry process at the authority's end \cite{NIDAonline}.

\subsection{SIM Card and Mobile Money Registration}
\label{SIM registration using IDs}
Although SIM card ID-verified registration in Africa started in 2006, the regulations were not strictly enforced until 2015 \cite{martin2021exclusion}.
In both Kenya and Tanzania, mobile subscribers are required to register new SIM cards using the stipulated authentication and registration procedures. Parents can use their IDs to register SIM cards on behalf of minor children who are further required to re-register these SIM cards under their own identification upon attaining the age of majority 
\cite{MinorSIMKenya} \cite{MinorSIMTz}. Tanzania has also resolved to issue IDs to under 18s \cite{MinorIDTz}. In Tanzania, the SIM card registration process entails the collection of fingerprints that are verified against those held on the national ID authorities' databases, to verify the ID is being used by the rightful person. Once a SIM card has been registered, users can activate their mobile money service, a process that requires more identification and verification. Figure \ref{figSIMRegistration} provides a general illustration of the two processes.

\section{Related Work}
The relevant prior work in this space can be divided into three categories: (1) SIM card registration processes (2) Challenges accessing ID in Africa and (3) Regulatory environment surrounding mobile money.

\subsection{SIM Card Registration and Implications}
Studies on SIM card registration have mainly focused on the advantages and disadvantages to both MNOs and mobile users, while a few have reported on privacy and security implications. 

Jentzsch highlights how registration, over time, leads to increased profits for MNOs, e.g., by enabling more targeted advertising\cite{jentzsch2012implications}. On the other hand, MNOs have been lost millions of dollars for failure to deactivate unregistered SIM cards \cite{martin2021exclusion}. 
For users, deactivation of SIM cards has led to increased usage of third-party SIM cards, especially among people with no ID \cite{stark2021mobile} \cite{martin2021exclusion} \cite{UNHCR}. The implications of third-party SIM card use have not been discussed outside of the humanitarian sector. UNHCR warns that the use of third-party SIM cards can facilitate the exploitation of displaced persons, as they may be required to pay unofficial fees for continued access to the SIM card \cite{UNHCR}. Other implications of SIM card deactivation were reported among Tanzania's urban poor by \cite{stark2021mobile}, who found that deactivation  can lead to loss of phone contacts. as the users often saved numbers to their SIM card.
This in turn could mean loss of employment opportunities, as people depend on referrals for day laborer opportunities from their social networks.

Jentzsch studied the privacy and security implications of SIM registration regulations, focusing on concerns over poor data protection regulations in African countries \cite{jentzsch2012implications}. The author observes that geographic patterns of users could reveal sensitive data such as their ethnicity and income levels. The author also found users in Mozambique often offered false or incomplete data during SIM card registration to protect their privacy. A study on users' perceptions on biometric SIM card registration in Bangladesh found similar opposition towards sharing fingerprints, which were viewed as personal data, and opposition towards the users' gender being identified, which could lead to harassment \cite{ahmed2017privacy}. The study also found SIM cards often changed hands with no formal records, as the local custom was to buy a SIM card along with the mobile handset, which undermined the purpose of registration for crime prevention. In contrast, similar studies in Nigeria \cite{oyediran2019attitude} and Uganda \cite{ssebakumba2018knowledge} had little mention of privacy. Users instead focused on the usefulness and ease of SIM card registration. 

Our work builds on prior work by focusing on the practices of third-party SIM card users and the consequences of using third-party SIM cards. 

\subsection{Challenges Accessing ID in Africa}
The 2021 ID4D Global Coverage Estimate report shows that in low income countries (LICs), 46\% of the unidentified say they do not have the required supporting application documents, 44\% live a long distance away from registration points, and 40\% feel the costs of obtaining an ID are prohibitive \cite{clark2022id4d}. Several studies have explored the origins of these challenges. UNICEF has reported on the unsupportive legal frameworks. Around half of SSA countries do not require birth registration within a month and more than half impose fees for birth registration, which can be a barrier to poor families \cite{unicef2017snapshot}. Even when registration is free, parents may incur costs for transport and for issuance of the certificate \cite{unicef2010strengthening}. Other studies have explored challenges faced by stateless and displaced persons \cite{maupeu2021kenya}, which include fleeing without supporting documents and regulations against double-registration. Double-registration restrictions have until recently restricted registered refugees from registering for national IDs. 

Those who are able to apply for IDs often face additional challenges. A report by the Center for Human Rights and Global Justice, Initiative for Social and Economic Rights, and Unwanted
Witness found non-indigenous Ugandans and ethnic minorities in Uganda face exclusion as the officials implemented a ``who you know'' basis for registration. Long queues at registration places and errors on IDs have also been reported  in both Rwanda and Uganda \cite{unwantedwitness} \cite{world2021people}. 

Voluntary exclusion from getting an ID is also common especially where there is distrust in the government or where alternative IDs are accepted. In Nigeria for instance, resistance towards national IDs is based on fears of government surveillance and authoritarianism \cite{okunoye2022mistrust}. Up until 2020, when national IDs became the only document accepted for SIM card registration, coverage of IDs in Tanzania languished at 24\% partly due to low demand. Our work complements these prior studies by exploring challenges around ID registration. We extend this knowledge by elucidating challenges at the different stages of the ID registration process. 

\subsection{Regulatory and Policy Environment around Mobile Money}
The growth, innovation, and adoption of mobile money can be attributed in part to the fact that governments have regulated it only lightly \cite{bahia2020} \cite{evans2014empirical}; however, this has come hand-in-hand with increased crime and fraud, forcing many countries to think more about the regulatory and policy environment surrounding the use of mobile money. Some of the regulations restrict the registration of SIM cards as discussed in Section \ref{SIM registration using IDs}  and require the use of identity documents to complete transactions. Some studies (e.g, \cite{ekow2022contribution} \cite{akomea2019control}) have focused on the impact of such legislation on mobile money adoption, showing that legislation influences adoption through the perceived trust that users establish.  In Uganda, contradictions among mobile money actors around legislation and directives on SIM registration created uncertainties for users about what was actually required \cite{kanobe2017policies}. While these studies elucidate the role of regulation in adoption and use of mobile money, they do not consider the practices of users who do not meet regulatory requirements (e.g., not having an ID to register a SIM card) or who adopt non-compliant practices (e.g., using a third-party SIM card). 

\section{Methods}
We conducted an exploratory qualitative study between July and August 2022. Participant recruitment and data collection was done with the help of in-country research assistants who worked for research consultancies that were competitively recruited prior to the study (Appendix \ref{app:contractorselection}). The studies were reviewed and approved by the institutional IRB and the respective national authorities (NACOSTI in Kenya and COSTECh in Tanzania).

\begin{table}
\centering
\begin{tabular}{l l l}
Participant Category &Kenya & Tanzania\\ 
\hline
1) Never registered a 1st party SIM card, only has a third-party SIM card & 7 & 6 \\ 

2) Tried but failed to register a 1st party SIM card, only has a third-party SIM card & 1 & 2 \\ 

3) Previously had a 1st party SIM card, only has a 3rd party SIM card now & 4 & 5 \\ 

4) Has both 1st and 3rd party SIM cards & 11 & 6 \\ 

5.a) Only uses a 1st party SIM card, Used 3rd party SIM card previously & 5 & 7 \\ 
5.b) Only uses a 1st party SIM card, Never used 3rd party SIM card & 8 & 10 \\ 
\end{tabular}
\caption{Number of participants for each category of SIM card users}
\label{table:participantcategories}
\end{table}

\subsection{Recruitment and Inclusion Criteria}
We recruited 72 adults (18 to 65 years old), 36 from rural areas and 36 from urban areas, who had a SIM card that was not shared with another person and that was used at least once a month. We included both third-party and first-party SIM card users to understand how their perceptions and experiences differed, and what factors enable people to overcome common challenges. These users were grouped further into five categories of SIM card users (Table \ref{table:participantcategories}).

A technical lead (one of the in-country research assistants) developed and oversaw the implementation of standard operating procedures for recruitment in both countries. We aimed to have an equal number of participants for each SIM category, gender, and age groups. In Tanzania, government officials (district and village officers and representatives of 10 households) suggested which streets to advertise on; snowball sampling was used for categories that were difficult to fill. The Kenya team primarily relied on snowball sampling and recruited on a rolling basis due to concerns the national elections ongoing at the time could lead to unrest. Interviews needed to be completed as soon as each participant was identified. Participants' demographics are summarized in Table \ref{tbl_demographics}.

\begin{table}
\centering
\begin{tabular}{l l l}
Gender & Kenya & Tanzania \\
\hline
Male & 20 & 18 \\
Female & 16 & 18 \\
\textbf{Age Range} & \textbf{Kenya} & \textbf{Tanzania} \\
\hline
18-25 years & 12 & 9 \\
26-39 years & 17 & 14 \\
40-49 years & 3 & 7 \\
50-65 years & 4 & 6 \\
\textbf{Literacy} & \textbf{Kenya} & \textbf{Tanzania} \\
\hline
Literate & 32 & 30 \\
Illiterate & 4 & 6 \\
\label{table:demographics}
\end{tabular}
\caption{Gender, Ages, and literacy levels of participants}
\label{tbl_demographics}
\end{table}

\subsection{Interview Guides and Interview Sessions}
We used structured interviews to ensure consistency given the multiple field assistants in the two countries. The questions covered the following topics (full questionnaire can be found in Appendix \ref{app:interviewguide}): 
\begin{enumerate}
    \item ID registration motivators, perceptions, and experiences 
    \item SIM card registration experiences 
    \item Perceptions of SIM card registration using IDs
    \item Accessibility and usage of mobile money savings, microloan, and insurance products 
\end{enumerate}

We started with questions on ID registration as that is the prerequisite for SIM registration. Questions on ID registration perceptions and experiences varied for different participants. We asked those who had never applied for an ID about their perceptions on the process and their perceived barriers that prevented them from applying. The remaining participants were asked about their ID registration journeys including the challenges faced, workarounds attempted, and proposed changes to the process. To avoid priming participants to focus on privacy and security, we asked about general consequences of using third-party SIMs.  

The interview guides were translated to the Tanzania and Kenya variants of Swahili by two of the researchers who are native speakers of Tanzanian and Kenyan Swahili. These translations were reviewed by other native speakers from each country (2 per country). 

We used cognitive walkthroughs to test the questions in both the English and Swahili guides for clarity and to identify biased and/or leading questions. We subsequently conducted 3 rounds of pilots, whose insights we used to revise and improve the interview guides. We did not include the pilot data in the final analysis. 

Interviews lasted approximately 75 minutes and included both this interview and a companion interview on mobile money practices, reported on in a separate paper. The SIM card interviews that we discuss in this paper were between 20 and 55 minutes long and occurred prior to the mobile money interviews. All interviews in Kenya were conducted face-to-face and approximately half of the participants chose to be interviewed in Swahili. All interviews in Tanzania were conducted over the phone and all participants chose to be interviewed in Swahili.  The variation in interview mode (face-to-face vs phone) was due to the different clearances provided by the respective national ethics review committees. All interviews were audio-recorded and transcribed to English by professional translation services and reviewed, through spot checks, by the two Swahili-speaking researchers.

\subsection{Participant Compensation}
Participants in Tanzania received the equivalent of \$8 (urban) and \$4, which are based on estimates for hourly wages in urban areas and daily work output in rural areas. The Kenyan national ethics committee felt the minimum wage estimates could be coercive and approved the equivalent of \$3.4 for both urban and rural participants.

\subsection{Data Analysis}
Two primary coders independently coded a random sample (10\%) of the data and then jointly developed a codebook. Single coding was used for the remainder of the data in an iterative manner, with spot checks conducted by the second reviewer at each iteration. Discrepancies that arose, mostly on whether to use parent codes or parent and children codes, were discussed and resolved. 

\subsection{Limitations}
While the use of structured interviews was useful in ensuring consistency, it may have imposed some limitations on the opportunistic data that could be gained through unstructured interviews.

A second limitation lies in the inherent constraints of qualitative work including the potential for over-reporting, under-reporting, and social desirability biases. The latter is particularly important given that we conducted the study in Kenya at a time when the regulators had issued a directive to all MNOs to re-register user SIM cards using biometrics \cite{Reporter_2022,Njeru_2022}. Thus, questions regarding third-party SIM cards may have elicited fear in participants imagining that any evidence of non-compliance would expose them to relevant authorities and attract undesired consequences. We did our best to mitigate for social desirability by emphasizing that our research was in no way connected to the directive, and that this was an independent study to understand the challenges that people face. We also reiterated that there were no wrong or right answers to our questions, but that we were interested in all user experiences. 

Finally, the results may be subject to recall bias as participants were expected to report details on the procedures, events, timing and costs of registration experiences.

\section{Results}
\label{Sec:Results}

\subsection{RQ1: Privacy and Security Implications of Using Third-Party SIMs}
We were interested to learn how people acquired third-party SIM cards as well as the challenges and risks that emerged in using them. We found 2 risky practices around third-party SIM registration, which we describe in \ref{sec:IDUseWithoutConsent} and \ref{sec:IgnoringConsequences}. We then discuss the 3 privacy and security implications of using third-party SIMs in \ref{sec:KYCcompliance}, \ref{sec:InsecureAccess}, and \ref{sec:MoMoprivacy}. 

\subsubsection{Using other people's IDs to register SIM cards without consent}
\label{sec:IDUseWithoutConsent}
Third-party SIM cards were acquired either through (1) social ties or (2) buying pre-registered SIM cards from agents. 

Most (n=16) of the 28 active and former third-party SIM card users from Kenya registered under family members' IDs --- usually parents, siblings, and other relatives. One participant even used a deceased relative's ID. The remaining registered under friends' or acquaintances' IDs. Tanzanian participants also mostly used family members' (n=11) or friends' (n=5) IDs. In both Kenya and Tanzania, agents also sometimes used their IDs to register for clients, but this service incurred additional costs. KE34 explained: 
\begin{quote}
''I pleaded with [the ID owner], and she said, `If it's that way, you'll have to pay me'. I asked her: `How much do you want me to pay you?' Honestly, that day, she requested 200 shillings [equivalent to \$1.47]. I gave it to her, and she gave me the ID.''
\end{quote} 

In most cases in Kenya, participants took the ID and did the SIM registration themselves.
Some agents attempted to verify the participants had permission to use the ID (n=8), but this was often limited to asking questions on the relationship between the ID owner and the participant and why the ID owner was unable to accompany the participant. As a result, two participants were able to use their relatives' IDs without consent. 
\begin{quote}
''At first, I asked my mother but she refused because she doesn’t want anyone to share with her ID. So, I went to her handbag, took it, and went to register with it.'' (kE15)
\end{quote}

In cases where the agent and client had a relationship, no checks were made. This resulted in one participant's mother (KE34) using his ID to register a ''questionable'' person without his consent. 

Since SIM registration in Tanzania requires biometrics collection, third-party SIM registration could, in theory, only be done in the presence of the ID owner. Despite this, we found some third-party SIM users (n=4) who had bought pre-registered SIM cards from agents. Interviews with first-party SIM users in Tanzania revealed agents would register multiple SIM cards using clients' IDs and biometrics, and keep the extra to sell later on.
\begin{quote}
''I was told [by the agent] that I have registered five times. I used my ID to register a SIM card for only one person - my child.'' (TZ11) 
\end{quote}

\subsubsection{ID owners registered third-party SIMs for others despite security concerns or risks}
\label{sec:IgnoringConsequences}
In both countries, participants who had registered SIM cards for others (n=6 in Kenya, n=3 in Tanzania) under their IDs explained that the beneficiary's lack of an ID coupled with their need for mobile services was their main motivator for doing so. Three participants from Kenya expressed concerns over the third-party user misusing the SIM, but decidedd to disregard them and assist the third-party user. These concerns might not be unfounded. One third-party SIM user (KE10) recounted using services like MoMo microloans on his third-party SIM, rather than his first-party SIM, as any negative consequences would not fall on him. He explained: ``If you borrow using another person’s SIM card, you do not care if they are listed in [the Credit Reference Bureau]....You can put them at risk but you do not care that much, as long as yours is safe.'' 

The remaining SIM benefactors expressed full trust in the beneficiaries. However, in two participants' cases, one Kenyan and one Tanzanian, trust was extended to people they had met only a few times.  

\subsubsection{Inability to comply with KYC requirements which led to risky workarounds}
\label{sec:KYCcompliance}
In Kenya, MoMo users have to present their ID to the agent when depositing or withdrawing money to/from their accounts. All participants therefore used workarounds to access agent services. These included: (1) borrowing the ID used to register the SIM card or memorising the ID details (2) establishing relationships with specific agents and, when travelling, transacting only with agents introduced to them by friends (3) using agents who don't usually ask for IDs (4) breaking down large transactions into multiple small ones because agents are less likely to ask for IDs for smaller transactions (4) asking others (sometimes strangers) to complete transactions on the participant's behalf. With regards to the latter,  KE36 said:
\begin{quote}
''You may be in a place [and you don't have] the physical ID...[So] you ask a person that you find at the shop: 'Do you have an ID? I don’t have mine right now...Can I send you the money then you withdraw it for me?''
\end{quote}

In both Kenya and Tanzania, replacing a SIM card after the original had been lost, damaged, or blocked also required presenting the ID used to register the old SIM card (and biometrics in Tanzania's case). Furthermore, communicating with customer service required the user to provide ID details. Third-party SIM users therefore frequently had to rely on the SIM benefactor to access various services. TZ34 remarked: 
\begin{quote}
''I was supposed to renew the SIM card so that I am able to use it again. But the person who helped me to register it is in Shinyanga and I am in Dar es Salaam [600 miles away] and it was a bit difficult. So, I asked her to renew it and then send it to me through a bus.''
\end{quote}

Participants expressed several concerns and experiences that stemmed from relying on others to access mobile services:

\subsubsection{Insecurity over maintaining access to services and subsequent financial losses.}
\label{sec:InsecureAccess}
Several participants (n=3 in Kenya, n=6 in Tanzania) expressed fears over being unable to reactivate a blocked SIM card or resetting a forgotten MoMo PIN, which can lead to temporary or permanent loss of savings. This could happen if the SIM benefactor exceeded the limit on number of SIM cards that can be registered under one ID, if they lost contact with the SIM benefactor, or because the benefactor had decided to deactivate the SIM card. The distrust existed even where the benefactor was a close relative e.g., a parent.
\begin{quote}
''I mostly do not keep money [in the third-party SIM] because I am not sure about it. If [my mother] says [the SIM] should be blocked, it will be a process (to regain access). I fear keeping money in it.'' (KE15)
\end{quote}

Some third-party SIM cards were also at risk of being deactivated. To comply with the Financial Action Task Force (FATF) recommendations on anti-money laundering and terrorist financing, MNOs in Kenya and Tanzania frequently conduct verification checks where SIM users are required to either input the ID number used to register the SIM card (via USSD) or visit the nearest agent to present the physical ID. Unverified SIM cards are de-registered (blocked). In both countries, the third-party SIM users who had bought pre-registered SIM cards or agent-registered SIM cards found it difficult to complete this step, as they had no contact with the ID owner. For some, like TZ24, this resulted in frequent de-activation. She remarked:
\begin{quote}
''I registered SIM cards five times. They deactivate them every time that. Some days, when I register in the morning, they deactivate it by evening. Sometimes, it takes a week. The longest I have had a SIM card was two months.''
\end{quote}

To avoid financial losses from the risk of not having access, participants reported voluntarily self-excluding from using MoMo for savings. 

\subsubsection{Reduced privacy over the MoMo account.}
\label{sec:MoMoprivacy}
Third-party SIM users felt vulnerable to the possibility of the SIM benefactor knowing their financial details, for example, when needing to replace their SIM card. Thus, to further enhance their privacy, most participants (n=10 in Kenya; n=6 in Tanzania) who had both a first- and third-party SIM indicated that they preferred using their personal SIM card for savings and credit services.  KE17 explained, ``I do not save in the [SIM card] with my husband’s name. Because I want it to be personal.'' 

\subsubsection{Mobile money frauds and potential implication in crimes.}
\label{sec:frauds}
In both countries, participants who had registered under agents or distant acquaintances' IDs feared being victims of SIM swap scams and losing their savings. 
\begin{quote}
''There is no way to protect [the savings]. It's not my ID. It's just that the owner has not decided to wrong me. [The MNO] can change the PIN if you have the ID.'' (KE35)
\end{quote}

A few participants (n=2 in Kenya, n=2 in Tanzania) expressed fears of being implicated in crimes committed by the SIM benefactor, either because of mistaken identity (law enforcement officials believing the person registered on the SIM is the same as the user) or because of mistaken association (officials believing the third-party SIM user must know the SIM benefactor). These were founded on personal and others' past experiences. According to TZ32: 
\begin{quote}
I had my friend whose SIM card [given to a third-party] was used in something like fraud and it was his name. So, he experienced a lot of challenges, taken to the police and such things.
\end{quote} 

\subsection{RQ2: Factors that Contribute to the Use of Third-Party SIM Cards}
We asked participants why they used third-party SIM cards. The reasons given fall into 2 groups: (1) Difficulties in getting an ID and (2) Non-ID related factors such as misconceptions and circumstances at the time of acquiring the SIM card. We elaborate on these two factors in Sections \ref{sec:IDrelatedfactors} and \ref{sec:NonIDrelatedfactors}.

\subsubsection{ID-Related Factors}
\label{sec:IDrelatedfactors}
Most participants (17 in Kenya, 21 in Tanzania) did not have an ID at the time of SIM card registration. In Kenya, this was usually because the participant had registered the SIM when under 18 years old, and therefore ineligible to apply for an ID. In Tanzania, participants without an ID at the time of SIM card registration were evenly split in 2 groups. The first group had applied for an ID but faced delays in receiving the ID number or physical card, which are needed for SIM registration. Meanwhile, the second group had chosen not to apply for an ID for reasons including: not being able to spare the time or energy needed for the process; and being put off by the convoluted process. 

\begin{figure*}
\includegraphics[width=\textwidth, height=10cm]{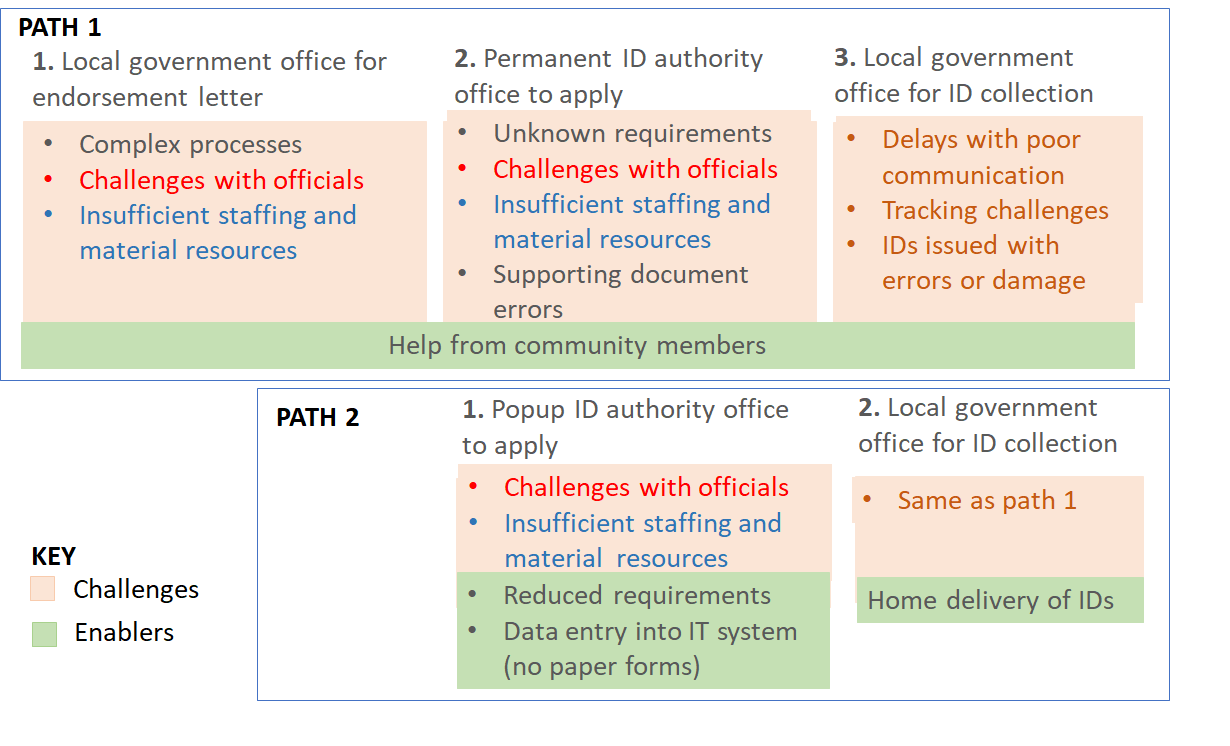}
\caption{Challenges and enablers at different stages of the ID registration process. Colored text is used for factors recurring in different stages. 
}
\label{fig:IDprocesschallenges}
\Description{A diagram summarizing the challenges and enablers experienced at different stages of the ID registration process. The challenges differ depending on whether the ID applicant registered at a fixed office - path 1 - or whether they registered at a popup registration point - path 2. Path 1 has 3 steps - getting an endorsement letter, applying for the ID, and collecting the ID. when getting an endorsement letter, applicants face 3 challenges: complex processes, challenges with officials, and insufficient staffing and material resources. At the ID application office, applicants face 4 challenges: Unknown requirements, challenges with officials, insufficient staffing and material resources, and supporting document errors. When collecting the ID at the registration office, applicants face 3 challenges: delays with poor communication, tracking challenges, and IDs issued with errors or damage. The main enabler at all 3 stages - endorsement, application, and ID collection - is support from community members. Path 2 has 2 steps: registering at a popup center and collecting the ID. At the popup ID registration point, applicants face 2 challenges: challenges with officials and insufficient staffing and material resources. The enablers for successful registration are reduced requirements and officials directly entering data into computer systems, which reduces the need for paper forms. At the ID collection step, applicants face 3 challenges which are the same as in path 1, namely: delays with poor communication, tracking challenges, and IDs issued with errors or damage. The enabler for fast delivery of IDs is home delivery by officials.}
\end{figure*}

First-party SIM card users indicated that challenges during ID registration were common, and there were multiple possible points of failure. Overall, more than 1 in 3 participants in both countries were unable to complete the ID registration process on their first visit to the registration authorities. Challenges revolved around: (1) complex processes and unknown requirements (2) insufficient staffing and materials in government offices (3) challenges associated with supporting documents (4) challenges interfacing with officials (5) long ID processing times with poor tracking and communication mechanisms and (6) ID processing errors. Figure \ref{fig:IDprocesschallenges} summarizes the challenges and enablers at different stages.
\begin{table}
\caption{ID registration attempts and success }
\label{ID-journey-table}
\begin{tabular}{lcc}
  & No. participants (Kenya) & No. participants (Tanzania)\\ 
 \hline
Successfully registered on first attempt & 18 & 19 \\ 
Successfully registered after 1 failed attempt & 7 & 5 \\ 
Successfully registered after multiple failed attempts & 5 & 4 \\  
Failed to complete registration & 1 & 2 \\ 
\hline
\end{tabular}
\end{table}

\paragraph{Complex Processes and Unknown Requirements. }
Participants were often unaware of process requirements (n=1 in Kenya, n=3 in Tanzania) including who is eligible for an ID, where to apply for the ID, and the needed information and supporting documents. Consequently, some failed to complete registration because they were ineligible or because they had incomplete documents.``The first day, since it was my first time, I didn’t know the process. I went, I inquired, and I took a letter from the chief but I didn’t take the original birth certificate'' (KE33).

Participant TZ28 highlighted how even officials sometimes disagreed on what supporting documents were acceptable, resulting in confusion for the applicant: 
\begin{quote}
They told me to go get an affidavit [of birth] and when I did, they told me that I have to go and change [my] whole name or rectify the whole birth certificate. When I went after a couple of weeks, the other [official] just took [the affidavit]. 
\end{quote} 

In both countries, some participants had to take additional steps to prove their ID eligibility due to a requirement to submit an endorsement letter from local government offices and to provide a detailed family history including parents' names and place of birth (district and constituency) and in Kenya - the clan name. This was required despite the applicants having all other necessary supporting documents. The participants most affected were those who were orphaned or from single-parent families, those from border towns, and those who had recently moved to the area. 
TZ16 explained:
\begin{quote}
A village executive officer or the chairman must recognize you as a resident of that area but it could be you have just moved in the area, so they cannot know all the residents, even the local representatives of ten houses couldn’t know about it. Therefore, you need to have a person who is well known, and go with them there so that they could say that you are really a resident in that area.
\end{quote} 
KE1, who didn't know his father also explained. ``I was advised to visit the chief’s office, get elders and get people who know me, people who can tell where I come from.''

\paragraph{Insufficient Resources. }
In both countries, material (paper forms, stamps, printing facilities) and human resources were often in short supply in both local government and ID registration offices (fixed and popup). To deal with insufficient material resources, some participants paid for the resources---sometimes at hiked up prices---while others had no option but to return on a different day. TZ18 remarked, ``[The officials] told us there were very few forms compared to the number of people so we contributed 200 shillings each.'' 

Human resource shortages were more challenging to navigate. In the worst cases reported by participants, we found one official was assigned to work between 2 village posts---which required traveling back and forth every few hours---and two officials were assigned to register around 1,000 people in a day. Unsurprisingly, most participants reported queuing for hours (usually 3--6 hours), or even days, at a time. To avoid long queues, some participants queued from the middle of the night, while others resorted to queue jumping, either through bribes or by requesting favors from the registration officials they knew. This could result in ''chaotic scenes''. In some locations, officials would skip form-filling and directly enter applicants' data into the IT system, which slightly eased the process. 

Queuing could have severe implications for the applicant and their families. Participants who were day laborers reported losing income on the days they went for ID registration. For instance, KE19, a farm worker, skipped work for 3 days. She said, ``When I used to go harvesting, I earned 150 shilling a day. So, for 3 days I lost 150 multiplied by 3 (\$3.34 over 3 days).''Meanwhile, stay-at-home parents reported their families going hungry as they were unable to leave the queue to cook. 

Tanzanian participants who used popup registration faced the additional pressure of being available to queue during a specific time period. Illness, work, and other time constraints prevented them from applying. A participant explained that popup registration drives are irregular and traveling to urban areas can be expensive and logistically challenging, especially if the applicant is unfamiliar with the area and needs someone to escort them. TZ1 explained:
\begin{quote}
[The registration place] is at Handeni town but I haven’t started going there. First, it requires you to go with an older person - a parent or someone who is familiar with that area. 
\end{quote}

Ultimately, participants frequently commented that queuing  was the ``worst part'' of the ID registration process. Some spoke of ``almost quitting'' and having to persevere because the ID was important to have, and the process might be even harder in the future. KE23 explained, ``If having an ID was not mandatory, I would have left it.''

\paragraph{Errors on Supporting Documents. }
During the document verification process (when getting the endorsement letter or submitting the ID application form), a few participants (n=4) found the wrong person named as a parent or the wrong birth date on their birth certificate, or a missing stamp on the school certificate. Changing the birth certificate details required visiting multiple offices. One participant, perhaps to avoid these processes, submitted the ID of the woman whose name was registered on her birth certificate instead of fixing the birth certificate errors. 

Tanzanian participants who applied in rural popup centers were spared these difficulties, as officials generally waived the requirement to present birth certificates or other supporting documents. 

\paragraph{Challenges with Officials. }
In both countries, absenteeism, tardiness, and sometimes rudeness complicated the process. In 2 cases, officials' chronic absenteeism resulted in participants being unable to continue the registration process. TZ25 quit because ``...it became a nuisance'' chasing after officials while KE24, who had been told to travel to his hometown at the border for an endorsement letter, never managed to find the right officials in the time he had: 
\begin{quote}
The junior staff I got there were unable to write me the letter, so they told me to wait for the senior person, but he is not available at the office, and he doesn’t come consistently during the week. I stayed there for two days, and there were things I needed to do in Nairobi so I had to go back. 
\end{quote}

When requesting help to understand the complicated process requirements, participants received little support and were sometimes even ``chased away.'' 

Officials also imposed unofficial charges for the material resources or for ``entrance/application'' costs. Unofficial ``expedited services'' were also offered. TZ10 explained, ``They said when I provide money, the identity will be given to me without any problem. They said it is the entrance fee.''  

\paragraph{Long ID Processing Times with Poor Tracking and Communication Mechanisms. }
In Kenya, several participants were given an expected wait time between 2 weeks and 3 months, before their IDs would be ready. Tanzanian participants were told they would be informed either through SMS or an announcement from the local government office. Five participants from rural Tanzania mentioned local government officials would post the full names along with photos or ID numbers of those whose IDs were ready for collection. 

Delays in ID processing were frustrating because participants' main motivator for applying for an ID was to access services or opportunities in the near future. In Kenya, the main opportunities were jobs (n=9), education (n=7), or registering a SIM card under personal name (n=5). In Tanzania, participants wanted to be able to: register a SIM card (n=4), travel (n=3), register for another ID e.g., entrepreneurship ID or driver's license (n=3), or to access loans (n=2). Frustrations were compounded by unrealistic expectations on processing time. TZ33 explained: 
\begin{quote}
What I expected was that, when I get there, I would complete all the information and obtain the ID right there because the machines were there and I believed they could do that. 
\end{quote}

One reason for delays in ID dissemination was the misrouting of ID cards to the wrong local government offices. 
KE 18 explained, ``You go where you applied for the ID, you find that it's not there, and they send you to another chief. They tell you to go and check there.'' 
When misrouting happened, community members stepped in to collect the lost IDs on others' behalf, as TZ11 explained:
\begin{quote}
I got my ID through a strange way. There are those who got their ID in the local government office, but I was at home and my younger sibling called me from Chanika town (158 miles away) and she asked me ‘sister have you received your ID’ and I told her ‘not, yet’. She told me there are a lot of IDs that have been poured down and I see your ID here.
\end{quote}

\paragraph{ID Processing Errors: }
Two Tanzanian participants revealed their IDs had been damaged or issued with the wrong details. For example, TZ12 explained, 
``I was told that my ID number was incorrect. Since the ID number was incorrect, I begun the whole process afresh.''

\subsubsection{Non-ID Related Factors}
\label{sec:NonIDrelatedfactors}
In Kenya, foreigners (n=3) used third-party SIM cards because of the erroneous belief that only national IDs were accepted during SIM card registration. In reality, regulations allow alien IDs and passports to be used for foreigners. Participant KE16's belief arose because agents had trouble registering SIM cards under his passport:
``Because I am a visitor here. So, I cannot get the ID to register the SIM card under my name.''
Participant KE15, a refugee, had no choice but to register for a third-party SIM card because SIM cards under refugee IDs are restricted from accessing MoMo services:  
``I wasn't happy to have no access to [MoMo] while other people do...That's when I looked for another person to assist me using their ID.'' 

On the other hand, citizens with IDs (n=4) got third-party SIM cards because of the desire for anonymity. In three cases, the participants had outstanding microloans from the MoMo provider and wished to avoid threatening calls and SMSs, and prevent the provider from automatically deducting from the deposits in their MoMo account to repay the loan. 
The remaining participant, KE1, found having a third-party SIM card helped him avoid ethnic profiling, especially in times of political uncertainty. He also did not want clients he transacted with to know his true identity and stalk him. He explained: 
\begin{quote}
Let’s say maybe you are in the street. Somebody has come and bought your stuff and their real motive is to get your personal number so they’ll say: 'I don’t have cash, can I send it (via MoMo)?' and because you don’t want money to go you tell them to send. But if you give them your real number…they get your Identity. 
\end{quote}
Lack of anonymity was also cited as a disadvantage by two other Kenyan participants. 

In Tanzania, participants had either bought pre-registered SIM cards at a time when national IDs were not required (n=2), had been handed down SIM cards by relatives (n=1), had to use a company SIM card (n=1), or found they had been mistakenly registered under another's ID (n=1). Unlike Kenyan participants, only 2 participants voluntarily switched to first-party SIM cards after regulations were introduced --- one after receiving his ID and the other after realizing his SIM card was not a first-party card. Other former third-party SIM card users only switched after the third-party SIM cards were lost, stolen, or blocked by the mobile operator (n=4) or after the the person under whose name the SIM card is registered (SIM benefactor) had died (n=1). As TZ10 explained: 
``(I registered a new SIM card because) I didn’t like the fact that her name (deceased sister) was showing up (during transactions). It made me recall memories.'' 

\subsection{RQ3: Perspectives on Mandatory SIM Card Registration Using Official IDs}
We asked participants about how they felt about the SIM card registration regulations. Our data revealed both the perceived advantages and disadvantages of these requirements. 

\subsubsection{Advantages of Registration Regulations}
Both first-party and third-party SIM card users in Kenya and Tanzania overwhelmingly supported SIM card registration because they felt it improves personal security and privacy, and eases access to services. They described how registration of SIM cards makes everyone identifiable, which makes it easier to track down fraudsters, but also makes people more conscious about how they use their mobile devices. KE10 explained: 

\begin{quote}
We can say [SIM card registration] helps because now when the swappers swap you, you can know the name of the person who has stolen from you. And also, you can know who just called you by sending money and reversing it.
\end{quote}

A few Kenyan participants also described how the requirement to present IDs during agent-facilitated transactions made phone sharing virtually impossible, which ensured personal privacy and greater security of MoMo savings. Meanwhile, participants from Tanzania noted that SIM card registration had the added advantage of increasing the demand for IDs among the general population. TZ6 explained,
``[People] now know the importance of IDs.... 
Because you can’t go to any office if you do not have a [national] ID'''

\subsubsection{Disadvantages of Registration Regulations}
Several participants (n=1 in Kenya, n=5 in Tanzania) felt SIM card registration could force people to use third-party SIM cards, thereby exposing them and the legal owners to greater risk. KE1 explained, ``We have children who are underage. Their parents registered SIM cards for them using their ID. 
Some [children] take loans and because it is under your ID, then the loan will retrace back to you.''  While participants specifically mentioned minors and orphans as at-risk groups, others, including TZ35, felt the wider population was at risk because of the restrictions on the type of IDs accepted for SIM registration and the bureaucracy of the national ID process: 
\begin{quote}
The youth are growing, the population is growing, if you have those (SIM card) restrictions, people will have to borrow other people's phones to make calls...It is easy to have a phone but hard to have a SIM card, so they should improve [the ID registration] process, so that a person is  able to register their SIM cards whether they have the [national] ID or any other ID.
\end{quote}

A few participants (n=4 in Kenya, n=2 in Tanzania) also spoke of the complexity introduced by other requirements like biometrics in SIM registration. KE20 noted, 
``The [facial biometrics collection] that was just introduced recently is a challenge. I don’t think the idea is good because someone like me, I have lost my ID and I cannot use the photocopy to get a SIM card.''  On the other hand, others felt SIM card registration without additional verification of identity e.g., through biometrics or stricter controls over agents, was insufficient to address security concerns.
\section{Discussion and Recommendations}
Kenya and Tanzania, like 155 other countries \cite{mobileidentity}, require physical identity proofing for pre-paid SIM card registration. Our findings show this is insufficient for proving the identity of the SIM card user, as third-party SIMs are easily acquired through social circles and through agents who intentionally register multiple SIM cards per ID to sell in the future. The use of 3rd party SIM cards further raises privacy and security implications for the user and SIM benefactor alike, and can also result in exclusion from using MoMo, the main driver for financial inclusion in SSA. While rethinking identity verification around SIM cards would be valuable, our results show that ID registration challenges contribute significantly to the the use of third party SIM cards. In this section, we discuss the insights that arise from our study and propose some recommendations that would help to address some issues that the study uncovered.

\paragraph{Finding 1: Agents are unable, and sometimes unwilling, to prevent unauthorized third-party SIM registration.} In some cases, they perpetuate the problem by using clients' IDs to register additional SIM cards for sale.
Both Kenyan and Tanzanian MNOs offer USSD or online services to verify the mobile numbers that have been registered uner a specific ID. However, given the potential for being implicated in crimes, protecting users from having their IDs being used without consent in the first place, would be preferable. Trends like eSIMs that replace the need for a physical SIM card have already been adopted by industry players like Apple \cite{sehgal2018esim}, and may impact the SIM registration landscape in Africa as the identity verification can be done by the user rather than the agent. A drawback of eSIMs is that they can only be deployed on smartphones and 46\% of mobile users in SSA still use basic or feature phones \cite{GSMAconnectivity}. Other agent-bypassing models would therefore be needed for this segment of the population. An alternative solution might be to use 2-step verification when more than 1 SIM card is registered to an ID. For instance, an SMS/USSD prompt can be sent to the first registered number to confirm their consent to register additional SIM cards.

\paragraph{Finding 2: The main driver for third-party SIM use is lack of first-party ID, largely due to unknown processes and insufficient human and material resources.}
Our results show that prolonged and repeated registration processes have disproportionately large economic costs for day laborers and those living in poverty, who comprise an increasing number of people in Africa \cite{worldbankpoverty}. Recurrent challenges can also lead to apathy and frustration. 

\paragraph{Finding 3: A desire for anonymity, especially when transacting, also motivates third-party SIM use.}
Our findings show that some clients desire anonymity to avoid ethnic profiling, stalking, or harassment. Mobile money providers typically employ SMSs that detail the transaction amount, date/time, sender or receiver name, and a reference number, to confirm successful transactions. This sensitive data can be accessed by unintended parties if the device is lost, stolen, or hacked. Tandon \emph{et al.} \cite{tandon2022know} found similar concerns with money transfer apps like Venmo in the US. They found that transaction memos contained a lot of data that compromised users' privacy. Several studies have explored privacy preservation in peer-to-peer mobile transactions \cite{paci2009privacy}  \cite{liu2023mobile} but their deployability in low resource environments (including a cost-benefit analysis) and usability are unknown. Research in this area is highly encouraged.

Based on these findings,  we recommend the following:

\textit{Recommendation 1: Leverage digital technologies coupled with human facilitators to plug gaps in the ID registration process.}
Kenya and Tanzania have launced digital platforms for online ID application and online or SMS-based tracking of issued IDs to address limited resources \cite{NIDAonline,KenyaIDTracking}. 
But the digital divide in the countries (only 51\% of people in Kenya and 22\% in Tanzania have a mobile phone with internet access \cite{malephane2022digital}) continues to persist.
MoMo agents have become a pillar to mobile money success in the region\cite{demirgucc2022global}. Such local community agents could be used to facilitate ID-related services. Rwanda for example has a network of 4,000 \textit{Irembo} kiosks---micro-businesses which help people complete online applications for civic services \cite{irembo} on the Irembo platform. The platform provides 104 e-government services and averages 1,500 daily users \cite{mutenyo2022digital}. However, porting this agent-driven approach elsewhere requires careful thought on managing incentive structures. Agents are often part of the local community but a number also roam from one neighborhood to another in search of clients \cite{stark2021mobile}. In Nigeria, agents conducting biometric SIM registration either deliberately flouted registration principles or refused to correct mistakes \cite{vanguardnigeria}. 

Alternatively, countries can explore roles that community leaders can play in streamlining ID registration processes 
Our findings revealed that community members' support (informational and practical) can be the differentiator between those who fail and those who succeed to navigate the ID registration process. Various African governments have local government units which oversee 10 households, villages or similar smaller clusters within the community \cite{were2021contextualizing}. Their geographical proximity and close social ties to citizens could be used to promote awareness on the importance of applying for IDs as soon as the person is of age, provide comprehensive knowledge on the registration process, and help facilitate the grievance, collection, and redress process.

\textit{Recommendation 2: Rethink identity proofing in the African context.} The majority of undocumented individuals live in Africa. Challenges proving eligibility for a national ID are thus likely to persist in the near future. India's Aadhar ID has decoupled nationality from identity and has also allowed ID owners to vouch for the undocumented. Both have allowed millions of undocumented people to access an ID. Similar systems can be trialled in Africa.

\textit{Recommendation 3: Governments and the United Nations should adjust ID access performance measurements to measure the number of people who successfully get IDs on the first attempt and within a reasonable cost and waiting time. }
Performance indicators for ID registration typically either focus on proportion of births registered (as with United Nations Sustainable Development Goal 16.9 \cite{statistics2019global}) or on the number of people who have successfully been issued an ID \cite{CAGTzNIDA}. We suggest that these indicators should also measure the cost to obtain ID, measured in units such as the number of tries and the time it takes individuals to obtain ID.


\section*{Acknowledgments}
The authors gratefully thank Michael Bridges for his valuable inputs on the design of this study. This study was made possible by the generous support of the Bill \& Melinda Gates Foundation. The views and opinions expressed in this study, however, are those of the authors and do not necessarily reflect the views or positions of the sponsors.

\bibliographystyle{plain}
\bibliography{references}
\appendix
\small
\begin{appendix}
\section{Interview Guide}
\label{app:interviewguide}
\textbf{ID Registration Experiences and Perceptions}
\begin{enumerate}
    \item Which ID do you need to register a SIM card in your name?
    \begin{itemize}
        \item For participants who tried and failed or who successfully completed ID registration 
        \begin{enumerate}
            \item When did you go to apply for an ID?
            \item Could you describe, step-by-step, everything that happened the last time you went to apply for an ID?
            \item How long did the entire process take (break down the time for transport, queuing etc.)?
            \item How much did the entire process cost (break down what items you paid for and how much each cost)?
            \item Were you unable to go to work or your business because of the registration process?
            \textit{If yes - Did you lose any income? If yes - how did you feel about this?}
            \item How did this experience compare to what you expected?
            \item What was the worst/hardest part?
            \item What was the best/easiest part?
            \item What was the process for getting a physical ID like? 
            \item How long did it take you to get the physical ID after you applied for the ID?
        \end{enumerate}
        \item For participants who have never applied for an ID
        \begin{enumerate}
            \item How do you feel about not having an ID?
            \item Why have you never tried to apply for an ID?
            \item What have you heard from others about the ID registration process?
            \item How have these stories influenced your attitudes to registering for an ID?
        \end{enumerate}
    \end{itemize}
\end{enumerate}
\textbf{SIM Card Registration Experiences}
    \begin{itemize}
        \item For participants with a third-party SIM card
        \begin{enumerate}
            \item When did you get this SIM card?
            \item How are you related to the person whose name is registered on the SIM card?
            \item Could you describe the steps you took to get this SIM card?
            \item How long did the entire process take?
            \item How much did the entire process cost?
            \item Did you experience any challenges when getting the SIM card?
            \item How did you overcome these challenges?
            \item Why did you get a SIM card under another person's ID?
            \item How do you feel about using a SIM card under another person's ID?
            \item What do you think about the requirement to have an ID in order to register a SIM card?
        \end{enumerate}
        \item For participants with a SIM under their own ID
        \begin{enumerate}
            \item When did you get this SIM card?
            \item Why did you choose to register the SIM card under your name?
            \item Can you tell me, step-by-step, everything that happened when you went to register the SIM card?
            \item How long did the entire process take?
            \item How much did the entire process cost?
            \item Have you ever used a SIM card under another person's name in the past? \textit{If yes - }
            \begin{enumerate}
                \item Why did you have a SIM card under another person's name?
                \item When did you stop using this SIM card?
                \item Why did you stop using this SIM card?
            \end{enumerate}
        \end{enumerate}
    \end{itemize}
\textbf{Experiences with Third-party SIM Cards}
\begin{enumerate}
    \item Do you use mobile money to keep your savings (yes/no)? \textit{If no - why not?; If yes - }
    \begin{enumerate}
        \item Do you put all your savings, around half, or less than half of your savings on mobile money? Why?
        \item Would you change how you think about or how you use mobile money savings if you had a SIM card under your name?
    \end{enumerate}
    \item Do you borrow on mobile money (yes/no)? \textit{If no - why not? If yes - }
    \begin{enumerate}
        \item Do the loans constitute all, around half, or less than half of your overall credit?
        \item Do you ever worry about being able to repay the loans? \textit{If yes - what is the worst that would happen if you were unable to repay?}
        \item Would you change how you think about or how you use mobile money microloans if you had a SIM card under your name?
    \end{enumerate}
    \item Do you use mobile money insurance products (yes/no)? \textit{If no - why not?}
    \item Are there services you have been unable to access or have had a lot of difficulty accessing because you didn’t have a SIM card registered in your name? \textit{If yes - }
    \begin{enumerate}
        \item Can you tell me about these incidents?
        \item Which of these services were you completely unable to access?
        \item Did you find some other way to access these services?
        \item Which of these services were you able to access but only after much difficulty?
    \end{enumerate}
\end{enumerate}

\section{Contractor Selection Process}
\begin {table*} [htbp]
\caption{Contractor selection process}
\centering
\begin {tabular}{|p{1.5cm}|p{6cm}|p{5cm}|p{2.5cm}|}
\hline
\textbf{Date}& \textbf{Activity}& \textbf{Outcome)}& \textbf{Supporting documentation}\\
 \hline
 \hline
31 Dec 2021& Advertising using a detailed Terms of Reference (TOR). The TOR was circulated with the help staff and students in our University. \medskip  & We received 8 applications & TOR and Statement of work (SOW)  \\
\hline
22 Jan 2022& PIs completed a preliminary review of applications to determine quality of submissions& There was too much variance in the quotes that we received. We therefore decided to clarify certain aspects of the project by sending a Statement of Work (SOW) to all applicants, and ask for a revised quote from all prospective contractors based on the SOW&   \\
\hline
11 Feb 2022& Following the revised applications, the PIs reviewed the submissions using the following criteria:
\begin{itemize}
\item Research Expertise
\item Ability to manage all countries
\item Asking price
\end{itemize}

This criteria had been discussed and agreed upon by the whole project team during the weekly meetings.
\medskip
& Three potential contractors were shortlisted &  \\
\hline
8 Mar 2022& Met each of the shortlisted contractors with the aim of understanding the following:
\begin{itemize}
\item Past clients and research engagements
\item Where the research work ended up (e.g., research outputs)
\item General proposed timelines
\item Budget and working flexibility
\item Existing team structure and presence in the countries of study
\medskip
\end{itemize} & We developed a comparative report comparing the three candidates & \\
\hline
16 Mar 2022& Part of the project team independently reviewed and ranked the three candidates. The PIs also shared the comparative report with the wider project team for their feedback and thoughts on the three contractors. \medskip  
& We selected 1 candidate that was the best fit  & \\
\hline
 \end{tabular}
\label{app:contractorselection}
\end{table*}

\end{appendix}

\end{document}